\documentclass[useAMS,usenatbib]{mn2e}         
\usepackage[dvips]{graphicx}
\usepackage{color}
\usepackage{amssymb,amsmath}

\begin{document}

\title[Abundances in intermediate-mass AGB stars]{Abundances in intermediate-mass AGB stars undergoing
third dredge-up and hot-bottom burning} 
\author[J.A. McSaveney et al.]
{J.A. McSaveney$^{1,2}$, P.R. Wood$^{1}$, M. Scholz$^{3}$, J.C. Lattanzio$^{2}$ and
K.H. Hinkle$^{4}$\\
$^{1}$Mount Stromlo Observatory, Research School of Astronomy and
Astrophysics, Australian National University, Cotter Road, \\
Weston Creek ACT 2611, Australia.  $^{2}$Centre for Stellar and Planetary Astrophysics, Department of
Mathematical Sciences,\\ Monash University, Melbourne, Australia.  $^3$ Institut f. Theoretische Astrophysik d. Univ. Heidelberg,
Albert-Ueberle-Strasse 2,\\
69120 Heidelberg, Germany, and School of Physics, University of Sydney, NSW 2600, Australia.
$^4$ National Optical Astronomy\\
 Observatory,  PO Box 26732, Tucson AZ 85726-6732, USA.
} 

\date{Accepted 2007 ?.
      Received 2007 ?;
      in original form 2007 ?}

\pagerange{\pageref{firstpage}--\pageref{lastpage}} \pubyear{2007}

\maketitle

\label{firstpage}

\begin{abstract}

High dispersion near-infrared spectra have been taken of seven highly-evolved,
variable, intermediate-mass (4-6 M$_{\odot}$) AGB stars in the LMC and SMC in
order to look for C, N and O variations that are expected to arise from third
dredge-up and hot-bottom burning.  The pulsation of the objects has been
modelled, yielding stellar masses, and spectral synthesis calculations have
been performed in order to derive abundances from the observed spectra.  For
two stars, abundances of C, N, O, Na, Al, Ti, Sc and Fe were derived and
compared with the abundances predicted by detailed AGB models.  Both stars
show very large N enhancements and C deficiencies.  These results provide the
first observational confirmation of the long-predicted production of primary
nitrogen by the combination of third dredge-up and hot-bottom burning in
intermediate-mass AGB stars.  It was not possible to derive abundances for the
remaining five stars: three were too cool to model, while another two had
strong shocks in their atmospheres which caused strong emission to fill the
line cores and made abundance determination impossible.  The latter occurrence
allows us to predict the pulsation phase interval during which observations
should be made if successful abundance analysis is to be possible.

\end{abstract}

\begin{keywords}
stars: AGB and post-AGB -- stars: abundances -- stars: oscillations.
\end{keywords}

\section{Introduction}

AGB stars are predicted to be major contributors to the enrichment of the
interstellar medium in carbon, nitrogen and s-process elements
\citep[e.g.][]{it78,dra03}.  The $^{12}$C and s-process elements are brought to
the surface of AGB stars during the third dredge-up \citep{ir83} and they are
then ejected into the interstellar medium by the extensive stellar winds that
occur during the AGB phase of evolution.  The $^{14}$N production is mainly
from the envelopes of the more massive AGB stars where the hot-bottom burning
process \citep{sdu75} can cycle the entire envelope through the outer parts of
the hydrogen-burning shell, converting $^{12}$C to $^{14}$N.  In the case where
the $^{12}$C in the envelope has a component that comes from third dredge-up, the
$^{14}$N resulting from the dredged-up $^{12}$C is of primary origin.  
There is strong evidence that some of the
$^{14}$N in the universe is of primary origin \citep{vl98}, with the primary
source dominating for metal abundances less than one third solar.  Hot-bottom
burning in intermediate-mass ($M \ga 3$M$_{\odot}$) AGB stars is a possible source for
this primary nitrogen.  Detailed stellar evolution and nucleosynthesis
calculations suggest that the hot-bottom burning in AGB stars may also involve
the Ne-Na and Mg-Al chains, leading to changes in the surface abundances of Al
amd Mg as well as nitrogen \citep{kl03,lw04}.  Unfortunately, all the evolution
and nucleosynthesis calculations are very uncertain due to the problem of
treating convection, especially the critical process of convective overshoot.
Observational constraints from intermediate-mass AGB stars are required to constrain the
convection parameters such as mixing length and overshoot distance, which are
critical to the nucleosynthesis results.  This study was initiated to provide
such constraints.

Considerable numbers of intermediate-mass AGB stars are known in the Magellanic Clouds
(Wood, Bessell \& Fox 1983), and these stars can be used to provide the
observational tests required for the evolution and nucleosynthesis
calculations.  Some work has been done in this area by \citet{smith2002} who
examined the C, N and O abundances in a sample of luminous AGB stars in the
LMC.  They found that $^{14}$N was enhanced in a way consistent with first
dredge-up only, without the need for hot-bottom burning.  However, the stars
studied by \citet{smith2002} were not known variable stars, meaning that they
were not near the end of their AGB evolution where the effects of dredge-up and
hot-bottom burning should be most pronounced.  In addition, the lack of
pulsation means that there was no way to estimate the mass of the stars
involved.

In this study, we have aimed to get surface abundances for a small sample of
intermediate-mass, pulsating AGB stars.  These stars are near the end of their AGB
evolution and are about to enter their final superwind stage where most of the
material currently in their envelopes will be ejected back into the
interstellar medium (in fact, two of the stars are IRAS sources already
exhibiting strong superwinds).  Since they are pulsating, we can also estimate
their current masses from pulsation theory.  On the other hand, because of
their low effective temperatures and their pulsation, the model atmospheres required
for the interpretation of their spectra are complicated and difficult, and must
involve the dynamics of the atmosphere.

In Section \ref{sec:observations}, we describe the sample of stars and the
photometry and spectra obtained for them.  In Section \ref{sec:pulsation}, we
describe the pulsation models, in Section \ref{sec:modelatmos} the model
atmospheres are described and the derived abundances are given.  The results
are discussed in the final section.

\section{Observations}\label{sec:observations}

\subsection{Near-infrared photometry}

The sample of objects observed in this study is given in Table \ref{tab:sample}.  All
these stars are oxygen-rich, pulsating, luminous AGB stars.  All the stars have
been monitored in $J$ and $K$ with the ANU 2.3m telescope at Siding Spring
Observatory, using either a single channel photometer (prior to March 1994) or the infrared
array CASPIR \citep{mcg94b}.
The mean bolometric
luminosity $M_{\rm bol}$ of each star was derived from the $JK$
photometry and monitoring, using the bolometric corrections from \citet{hou00a,hou00b} 
with $[Fe/H] = -0.3$ and -0.6 for the LMC and SMC, respectively.
Distance modulii of 18.54 and 18.93 and reddenings E(B-V) of 0.08 and 0.12 were
assumed for the LMC and SMC, respectively (Keller and Wood 2006).

The least evolved of the stars in Table \ref{tab:sample}, with lowest
luminosity, smallest amplitude and shortest pulsation period, is the most
luminous red giant in the LMC cluster NGC\,1866.  At the other extreme are two
IRAS sources with very long pulsation periods and large amplitudes.  The
luminosities and periods of the stars in the sample mean that their current
masses lie in the range 3--8 M$_{\odot}$ \citep{wbf83}.

\begin{table}
\caption{The intermediate-mass AGB star sample}
\label{tab:sample}
\begin{tabular}{lccccc}
\hline
Star             & $M_{\rm bol}$ & $P^1$ & $\Delta K$ & MC & Date$^{2}$ \\
\hline
NGC\,1866 \#4    & -6.00     & ~158 	 & 0.10       & LMC    &  030210 \\
HV 2576          & -6.61     & ~530      & 0.25       & LMC    &  030209 \\
HV 11303         & -5.59     & ~534      & 0.70       & SMC    &  030728 \\
HV 12149         & -6.84     & ~742 	 & 0.70       & SMC    &  020920 \\
GM 103           & -6.80     & 1070      & 1.30       & SMC    &  030728 \\
IRAS\,04516-6902 & -6.80     & 1090 	 & 1.30       & LMC    &  030211 \\
IRAS\,04509-6922 & -7.33     & 1290 	 & 1.50       & LMC    &  021202 \\
\hline
\end{tabular}
\newline\footnotesize{$^{1}$ Pulsation period in days.\\
$^{2}$ Date of Gemini observation in the form yymmdd.  For GM 103 and HV 11303
the observations were spread over 6 days and the given date is the mean value.}
\end{table}

\subsection{The high-resolution near-infrared spectra}

\begin{figure*}
\includegraphics[bb=84 225 514 565, width=13.5cm]{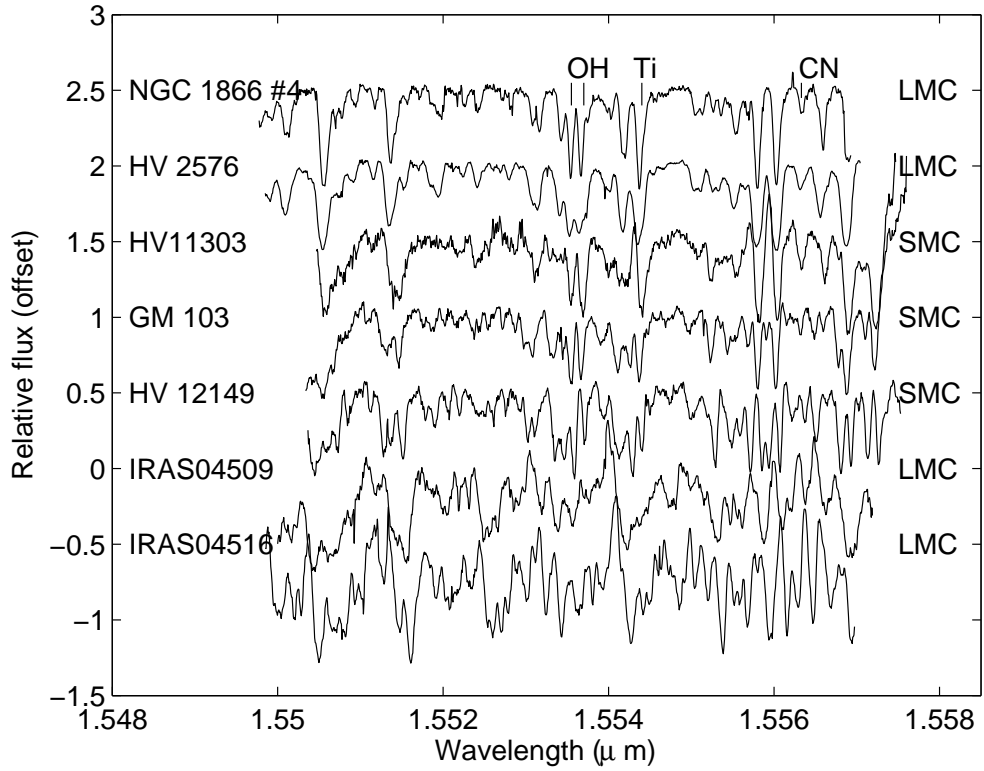}
\caption{The spectra of all objects in the region near 1.554 $\mu$m.  The
most important lines used for abundance determination are marked.}
\label{fig:h6420}
\end{figure*}

\begin{figure*}
\includegraphics[bb=84 225 514 565, width=13.5cm]{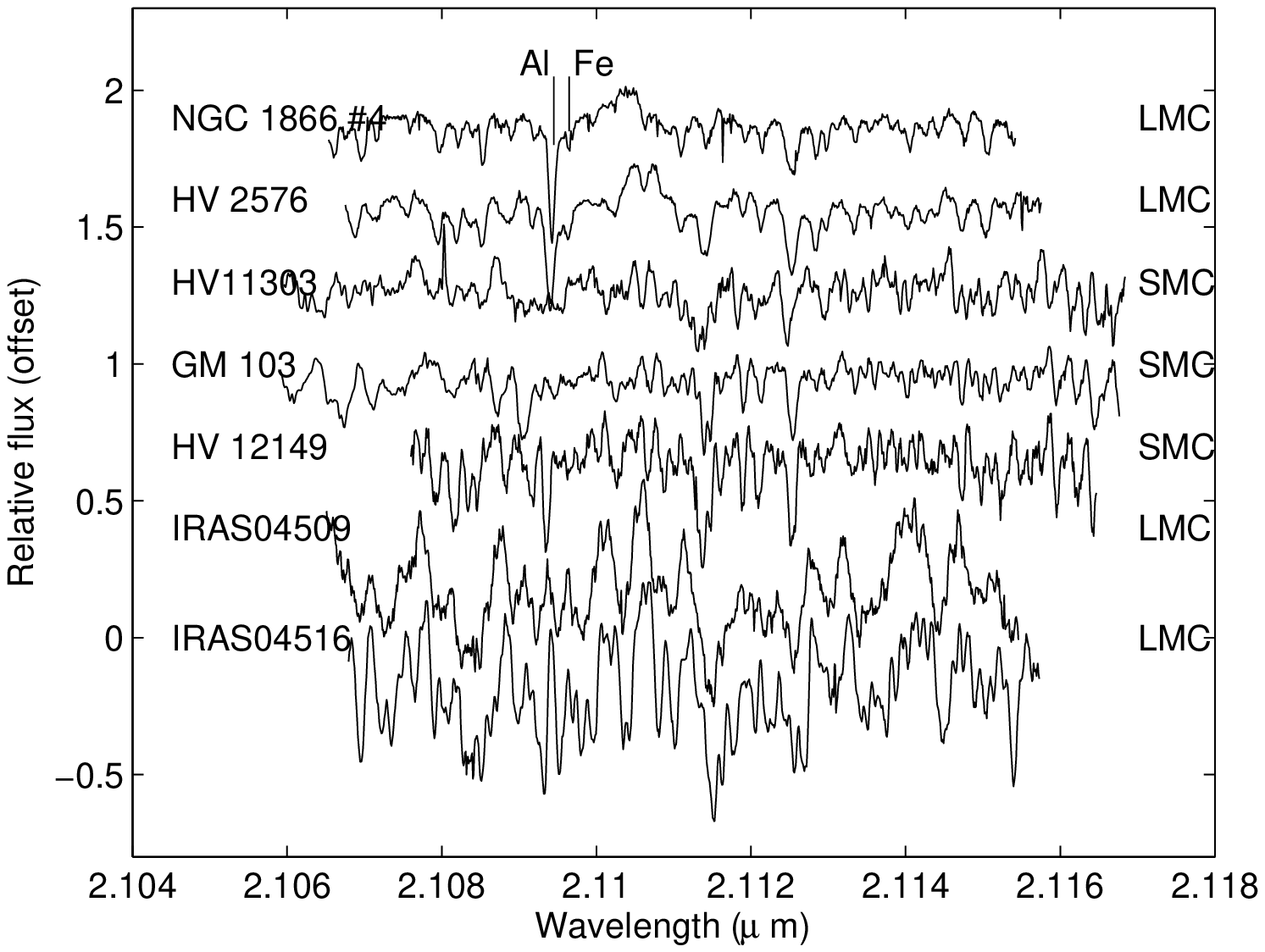}
\caption{The same as Figure~\ref{fig:h6420} but for the region near 2.112 $\mu$m.}
\label{fig:k4748}
\end{figure*}

\begin{figure*}
\includegraphics[bb=84 225 514 565, width=13.5cm]{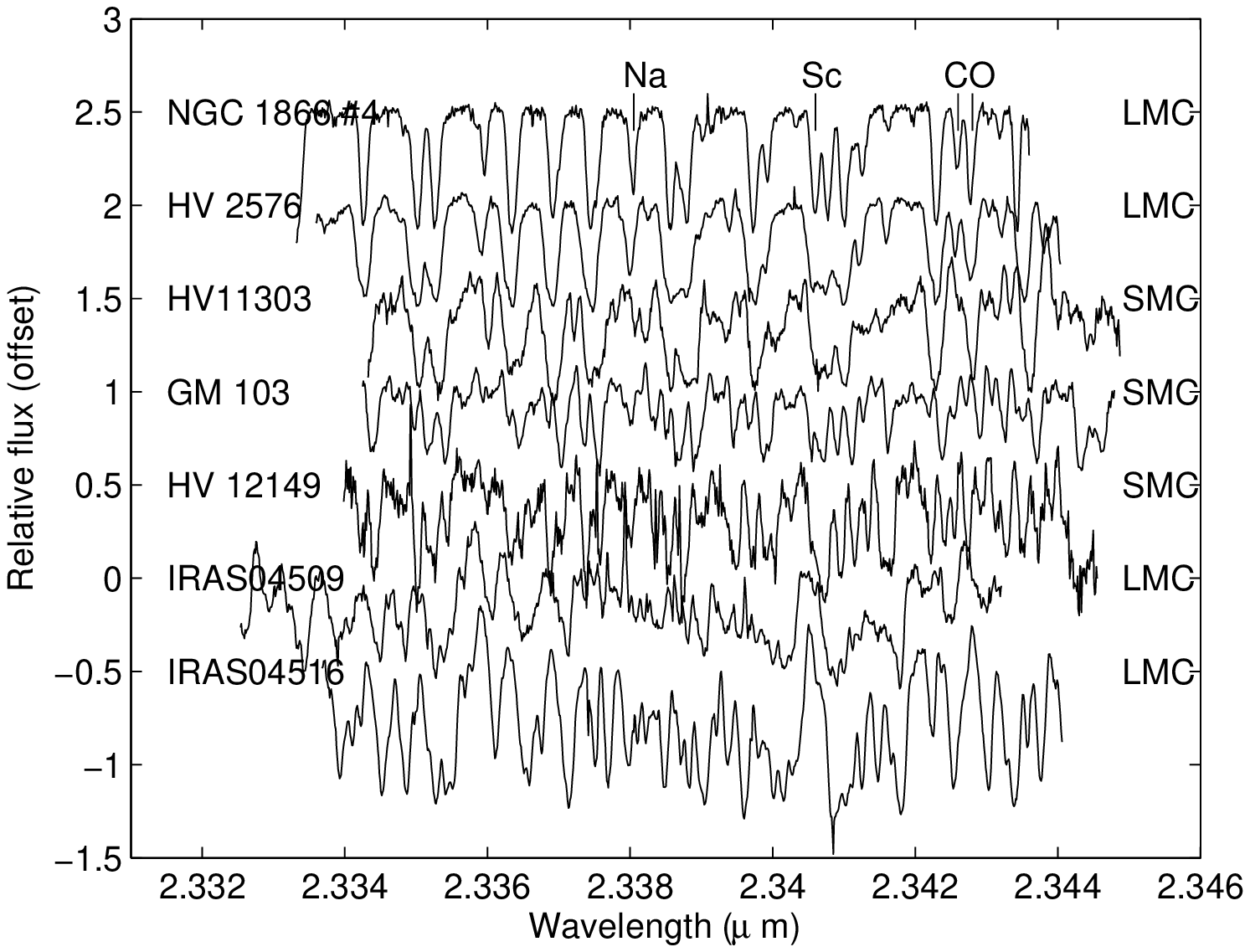}
\caption{The same as Figure~\ref{fig:h6420} but for the region near 2.340 $\mu$m.}
\label{fig:k4308}
\end{figure*}

Spectra were taken of these stars using the Phoenix spectrograph (Hinkle et al. 2003) on the
Gemini South telescope.  The observations were done in service mode, and the
observation dates are listed in Table \ref{tab:sample}.  For each star,
observations were made in 3 separate bands, as shown in Figures~\ref{fig:h6420},
\ref{fig:k4748} and \ref{fig:k4308}.  Firstly, an observation centred near
1.554 $\mu$m was made so that the OH lines could be measured for the
derivation of an oxygen abundance; lines of CN also occur in this piece of
spectrum.  Secondly, an observation near 2.340 $\mu$m was made to measure
$^{12}$CO (and possible $^{13}$CO) lines and hence the $^{12}$C abundance;
given the C abundance, the CN lines near 1.554 $\mu$m then yield a $^{14}$N
abundance.  The Na abundance was also obtained from this spectral region.  
Thirdly, an observation at 2.112 $\mu$m was made to include lines
of Al (and possible Mg).  Lines of Fe, Sc and Ti also occur in the
spectra, giving an estimate of the metal abundance.

\section{Pulsation models}\label{sec:pulsation}

Pulsation models and model atmospheres were made for four stars in
Table~\ref{tab:puls_mods}: NGC\,1866\, \#4, HV\,2576, HV\,11303 and GM\,103.
The spectra of the remaining stars in Table \ref{tab:sample} (HV\,12149, IRAS\,04516-6902 
and IRAS\,04509-6922) were considered too difficult to
model at this stage because of the extreme appearance of the spectra, with no clear continuum remaining
(see Figs.~\ref{fig:h6420}, \ref{fig:k4748} and \ref{fig:k4308}).  
This is presumably a result of the very large amplitude and very low effective temperatures
of these stars.   
The $J$ and $K$ photometry for the four stars for which modelling was attempted is given
in Table \ref{tab:jk}. Typical photometric errors are less than 0.03 magnitudes.  The $J$ and $K$
magnitudes have been converted to those on the AAO system of
\citet{ac83} using the conversions in \citet{mcg94a}.

\begin{table}
\caption{The modelled stars}
\begin{tabular}{lccccc}
\hline
Star          &  L/L$_{\odot}$ &  T$_{\rm eff}$ & M/M$_{\odot}$   & $\ell$/H$_{\rm p}$ & $\alpha_{\nu}^1$ \\ 
\hline	      	      		  		  	                            
NGC\,1866\,\#4 &  20000       	&  3490   	 & 4.0 		   &    2.57 	    &   0.05 	       \\
HV 2576       &  35000       	&  3350   	 & 6.0 		   &    2.21 	    &   0.51 	       \\
HV 11303      &  34000       	&  3490   	 & 4.6 		   &    2.21 	    &   0.58 	       \\
GM 103        &  41840       	&  3040   	 & 6.0 		   &    1.40   	    &   1.30 	       \\
\hline
\end{tabular}
\newline\footnotesize{$^{1}$ Turbulent viscosity parameter.}
\label{tab:puls_mods}
\end{table}

\begin{table}
\label{tab:jk}
\caption{$JK$ photometry}
\begin{tabular}{lccclcc}
\hline
JD24$^1$      & $J$-$K$ &  $K$           & & JD24      & $J$-$K$ &  $K$  \\
\hline
\multicolumn{3}{l}{NGC\,1866\,\#4}   & & \multicolumn{3}{l}{HV\,11303 (continued)}\\
50731 	  & 1.23    & 9.71    	     & & 52804 	   & 1.29    &	     10.02 \\
50772 	  & 1.27    & 9.69    	     & & 52924 	   & 1.26    &	     9.97  \\
50885 	  & 1.24    & 9.68    	     & & 53516 	   & 1.20    &	     9.71  \\
51157 	  & 1.30    & 9.67    	     & & 53574 	   & 1.20    &	     9.38  \\
52564 	  & 1.29    & 9.71    	     & &           &         &             \\
52714 	  & 1.33    & 9.64    	     & & \multicolumn{3}{l}{GM\,103}       \\
52803 	  & 1.25    & 9.69    	     & & 45980     & 1.66    &	     9.55  \\
52924 	  & 1.28    & 9.69    	     & & 46341     & 1.49    &	     8.65  \\
      	  &   	    &	      	     & & 46645     & 1.54    &	     8.57  \\
\multicolumn{3}{l}{HV\,2576$^2$}     & & 48142     & 1.63    &	     9.31  \\
52563	  & 1.39    & 9.00           & & 48168     & 1.70    &	     9.23  \\
52715	  & 1.43    & 9.13	     & & 48640     & 1.64    &	     8.36  \\
52804	  & 1.29    & 8.97	     & & 48851     & 1.62    &	     8.78  \\ 
52924	  & 1.36    & 8.87	     & & 48934     & 1.63    &	     9.13  \\
     	  &         &                & & 48992     & 1.67    &	     9.33  \\
\multicolumn{3}{l}{HV\,11303$^{2,3}$}& & 49259     & 1.89    &	     9.80  \\
49259 	  & 1.19    &	     9.81  & &   49317     & 1.87    &	     9.82  \\
49291 	  & -       &	     9.61  & &	 49374     & 1.68    &	     9.45  \\
49317 	  & 1.17    &	     9.47  & &	 52564     & 1.65    &	     9.33  \\
49374 	  & 1.19    &	     9.32  & &	 52714     & 1.59    &	     8.74  \\
52564 	  & 1.21    &	     9.27  & &	 52803     & 1.61    &	     8.61  \\
52715 	  & 1.39    &	     9.56  & &	 52924     & 1.64    &	     8.40  \\
\hline									   
\end{tabular}								   
\newline\footnotesize{$^1$ JD24 is Julian Date - 2400000.  $^{2}$ Extra photometry is given in Wood et al. (1983). $^3$ Extra
photometry is given in Catchpole and Feast (1981).}
\end{table}

For each observed star, given estimates of the mean luminosity $L$ and the
effective temperature $T_{\rm eff}$ from the $J$ and $K$ photometry, the
stellar radius was computed from the definition $L=4\pi\sigma R^2 T_{\rm
eff}^4$.  Then, given the radius and the known pulsation period, the current
stellar mass was computed from the $P$-$M$-$R$ relation of stellar pulsation.
These values of mass and luminosity were then used in the nonlinear pulsation
calculations for the star.

The nonlinear pulsation code described in Keller and Wood (2006), and
references therein, was used in this study.  The turbulent viscosity parameter
$\alpha_{\nu}$ was adjusted to give the observed $K$ light curve amplitude.
The $K$ and $V$ magnitudes were computed using $L$ and $T_{\rm eff}$ from the
pulsation models together with the bolometric corrections from
\citet{hou00a,hou00b}.

For three of the four modelled stars, MACHO light curves are available: the V
magnitude was computed from the MACHO magnitude using the transforms in Bessell
and Germany (1999).  For the fourth star, HV\,11303, an Eros2 light curve was kindly provided by
Patrick Tisserand (private communication), who also provided a conversion from the two
Eros2 bands to the V band.  The combined $V$ and $K$ light curves give a $V$-$K$
colour curve.  The effective temperatures of the models were adjusted (by
altering the ratio of mixing-length to pressure scale height $\ell$/H$_{\rm p}$
to reproduce the observed $V$-$K$ colour when available, or the $J$-$K$ in the
one remaining case of HV\,11303 (the Eros2 light curve was obtained after the modelling
was completed).  The final model parameters are given in
Table \ref{tab:puls_mods}.  In all cases, a helium abundance Y = 0.30 was assumed,
while a metal abundance Z=0.01 was assumed for the LMC stars and Z=0.004 was
assumed for the SMC stars.

In order to construct a model atmosphere taking account of the dynamical
processes in the outer layers, procedures similar to those described in
\citet{bsw96}, \citet{hsw98}, \citet{sw00} and \citet{ire04a,ire04b} were used,
with updates to the molecular opacity from \citet{ire06}.  Briefly, a pulsation
model at the same pulsation phase as the phase of the Gemini observation was
extracted from each pulsation series.  The radial profiles of gas pressure and
velocity for this model were transferred to a non-grey, spherical atmosphere code, and the
temperature structure was iterated to a new state, now consistent with the
atmosphere code.  Once the temperature, gas pressure, density and velocity
structure were all consistent in the atmosphere code, the model atmosphere
structure was transferred to a separate spectrum synthesis code which was used
to compute line profiles (see Section~\ref{sec:modelatmos}).

In the atmosphere code, half- and quarter-solar metal abundances
from \citet{gre96} were adopted for LMC and SMC stars, respectively.  Radiative and local
thermodynamic equilibrium was adopted and, in particular, the shock-heated
region behind the shock front was assumed to have negligible width and
to have no effects on the temperature stratification.  

The pulsation series and the extracted dynamical model used as input to the
atmosphere code is now described for each star.

\subsection{NGC\,1866 \#4}
\label{subsec:ngc18664}

NGC\,1866 \#4 is the most luminous red giant in the LMC cluster NGC\,1866 (the
designation \#4 is that given by Frogel, Mould and Blanco 1990: note that
Maceroni et al. 2002 have mislabelled \#4 as \#2 in their paper).  The MACHO
light curve of this star shows a period of 158 days.  Modelling the cluster
HR-diagram, \citet{bro03} derive masses for the He-burning giants of
3.8--4.15 M$_{\odot}$.  Our derived mass for NGC\,1866 \#4 is 4 M$_{\odot}$.
With this mass and the computed mean luminosity and $T_{\rm eff}$, this star
must be pulsating in the first overtone mode, a result also consistent with the
small amplitude.

\begin{figure}
\includegraphics[width=0.45\textwidth]{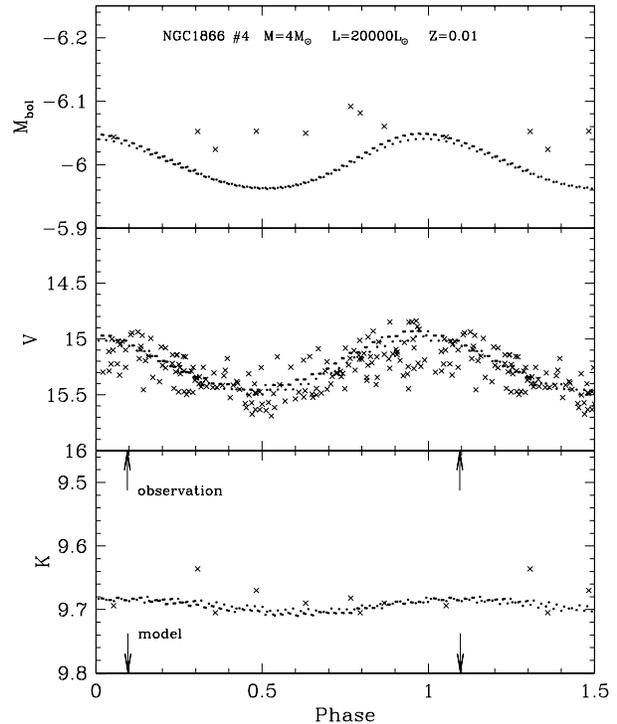}
\caption{The $K$, $V$ and $M_{\rm bol}$ light curves of NGC\,1866 \#4.  The
solid points are model values (several cycles are overplotted), while the crosses are
observations or $M_{\rm bol}$ values computed from the $J$ and $K$ photometry in
Table \ref{tab:jk}.  The $V$ values are computed from the MACHO $M_B$ and $M_R$ photometry
using the transforms of Bessell and Germany (1999).  The arrows show the phase of
the Gemini observations and the phase of the pulsation model extracted for model
atmosphere computation.} 
\label{fig:1866_lc}
\end{figure}

\begin{figure}
\includegraphics[width=0.45\textwidth]{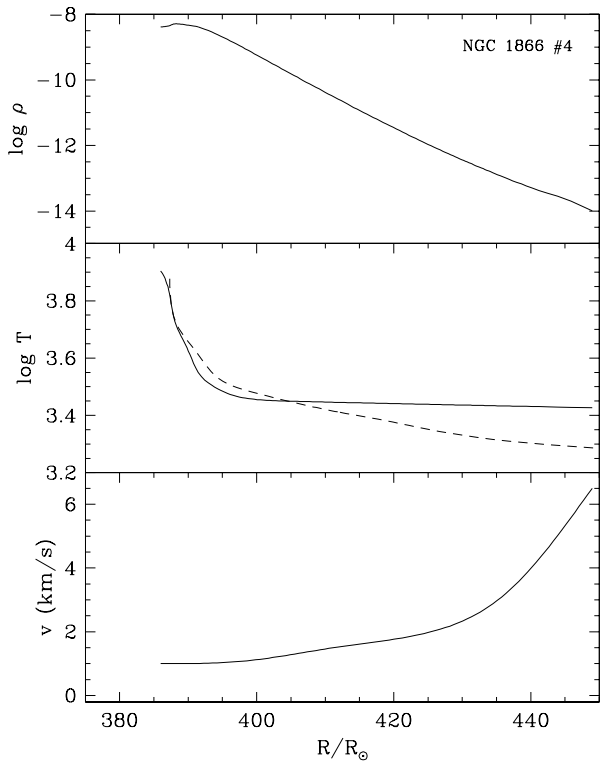}
\caption{The density $\rho$, temperature $T$ and velocity $v$ plotted against radius $R$ in a pulsation model
of NGC\,1866 \#4 at the same phase as the phase of Gemini observation of the star.  The dashed
temperature profile is that obtained from the non-grey model atmosphere code.} 
\label{fig:1866_atm}
\end{figure}

Fig. \ref{fig:1866_lc} shows the observed and computed $K$, $V$ and $M_{\rm bol}$
light curves.  The period and amplitude of the pulsation and the $V$-$K$ colour
are reproduced well.  The Gemini observation was made at a phase of $\sim$0.1
after visual maximum.  The structure of the pulsation model 
at this time is shown in Fig. \ref{fig:1866_atm}.  The
pulsation amplitude of this star is quite small so there are no shock waves
present in the atmosphere.  The temperature solution from the model atmosphere
program produces a cooling of the outer layers and a warming at moderate
optical depths compared to the modified Eddington approximation used in the
pulsation models.

\begin{figure}
\includegraphics[width=0.45\textwidth]{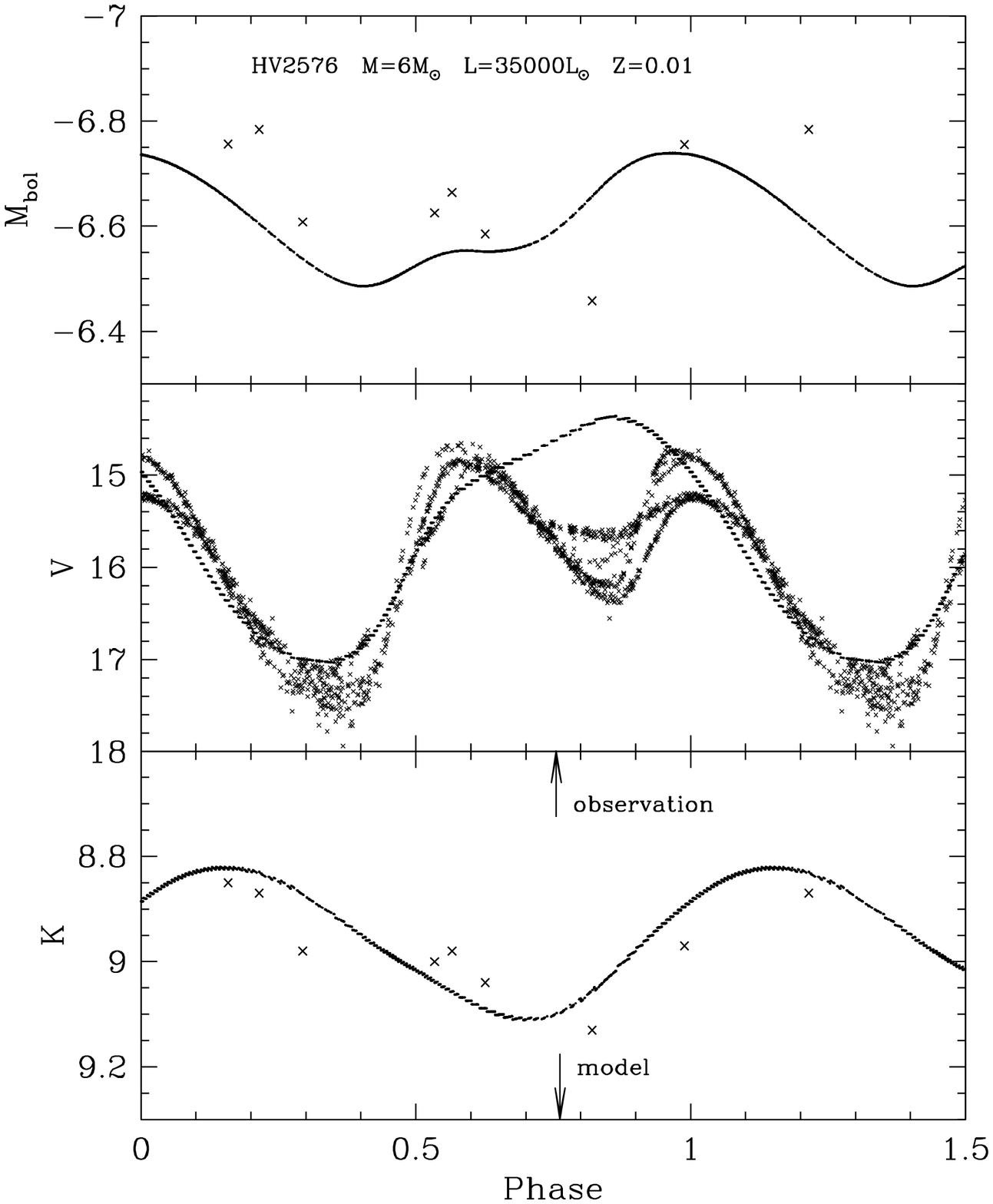}
\caption{The same as Fig.~\ref{fig:1866_lc} but for HV\,2576.} 
\label{fig:2576_lc}
\end{figure}

\begin{figure}
\includegraphics[width=0.45\textwidth]{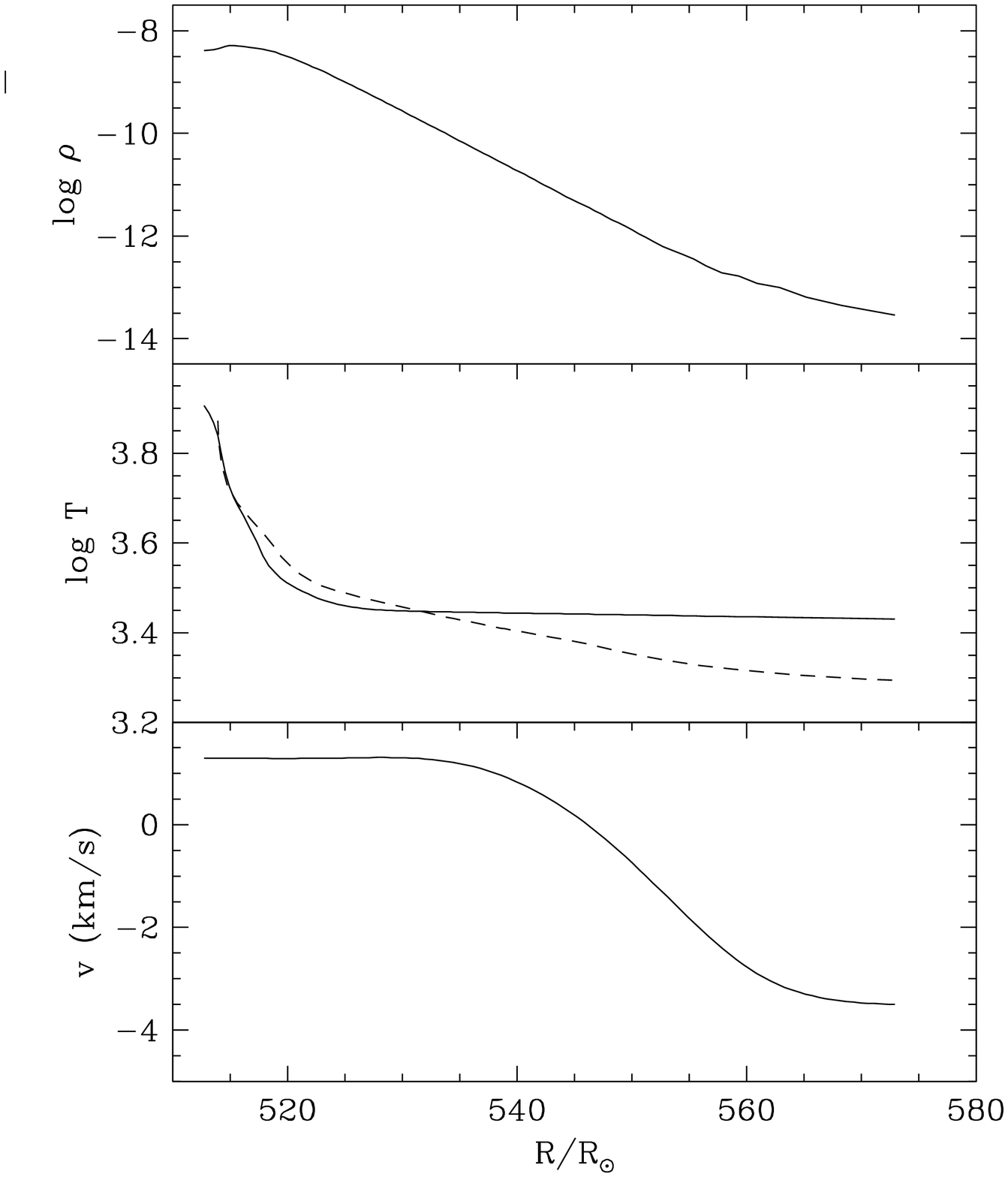}
\caption{The same as Fig.~\ref{fig:1866_atm} but for HV\,2576.} 
\label{fig:2576_atm}
\end{figure}

\begin{figure}
\includegraphics[width=0.45\textwidth]{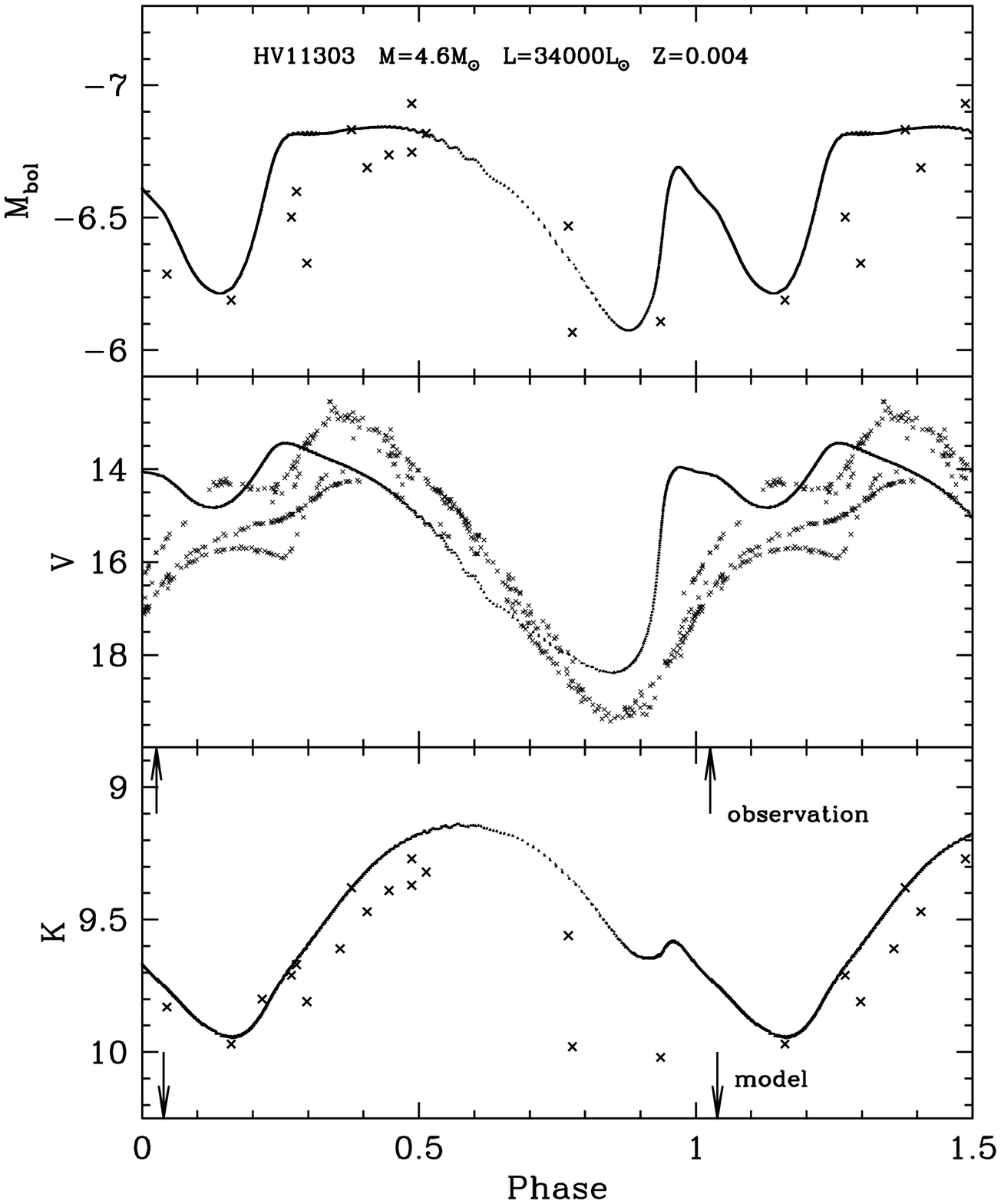}
\caption{The same as Fig. \ref{fig:1866_lc} but for HV\,11303.} 
\label{fig:11303_lc}
\end{figure}

\begin{figure}
\includegraphics[width=0.45\textwidth]{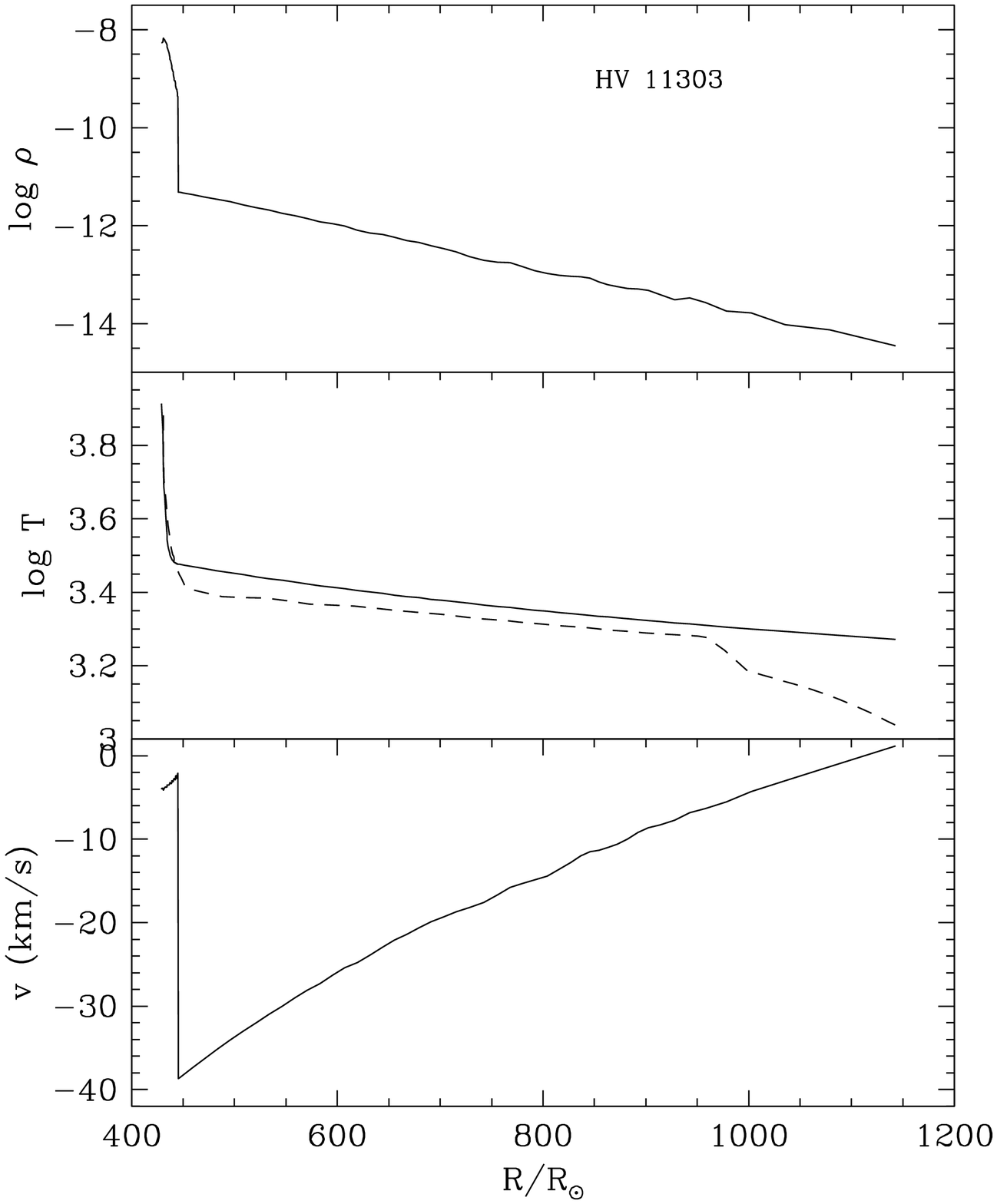}
\caption{The same as Fig. \ref{fig:1866_atm} but for HV\,11303.} 
\label{fig:11303_atm}
\end{figure}

\begin{figure}
\includegraphics[width=0.45\textwidth]{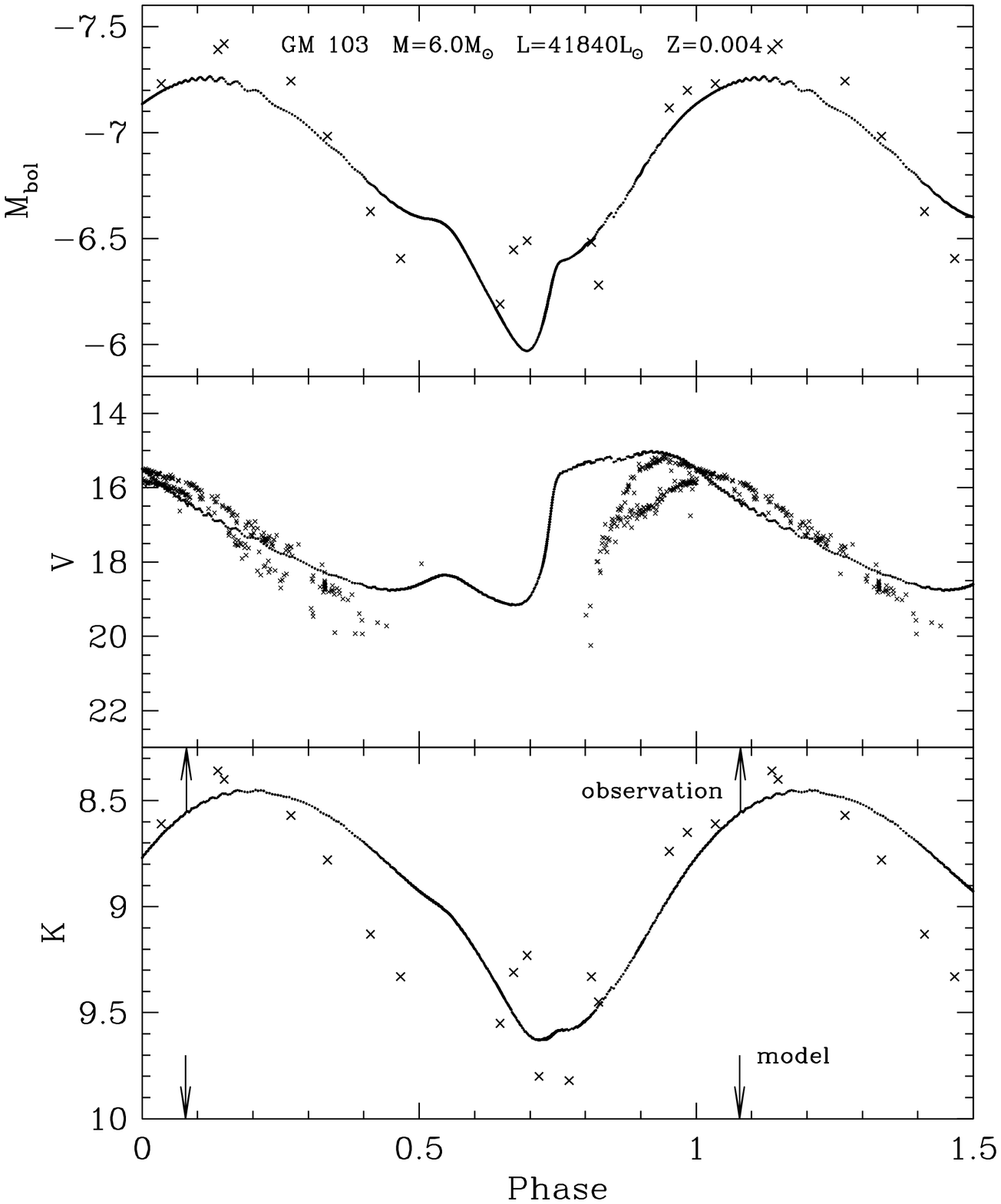}
\caption{The same as Fig.~\ref{fig:1866_lc} but for GM\,103.} 
\label{fig:103_lc}
\end{figure}

\begin{figure}
\includegraphics[width=0.45\textwidth]{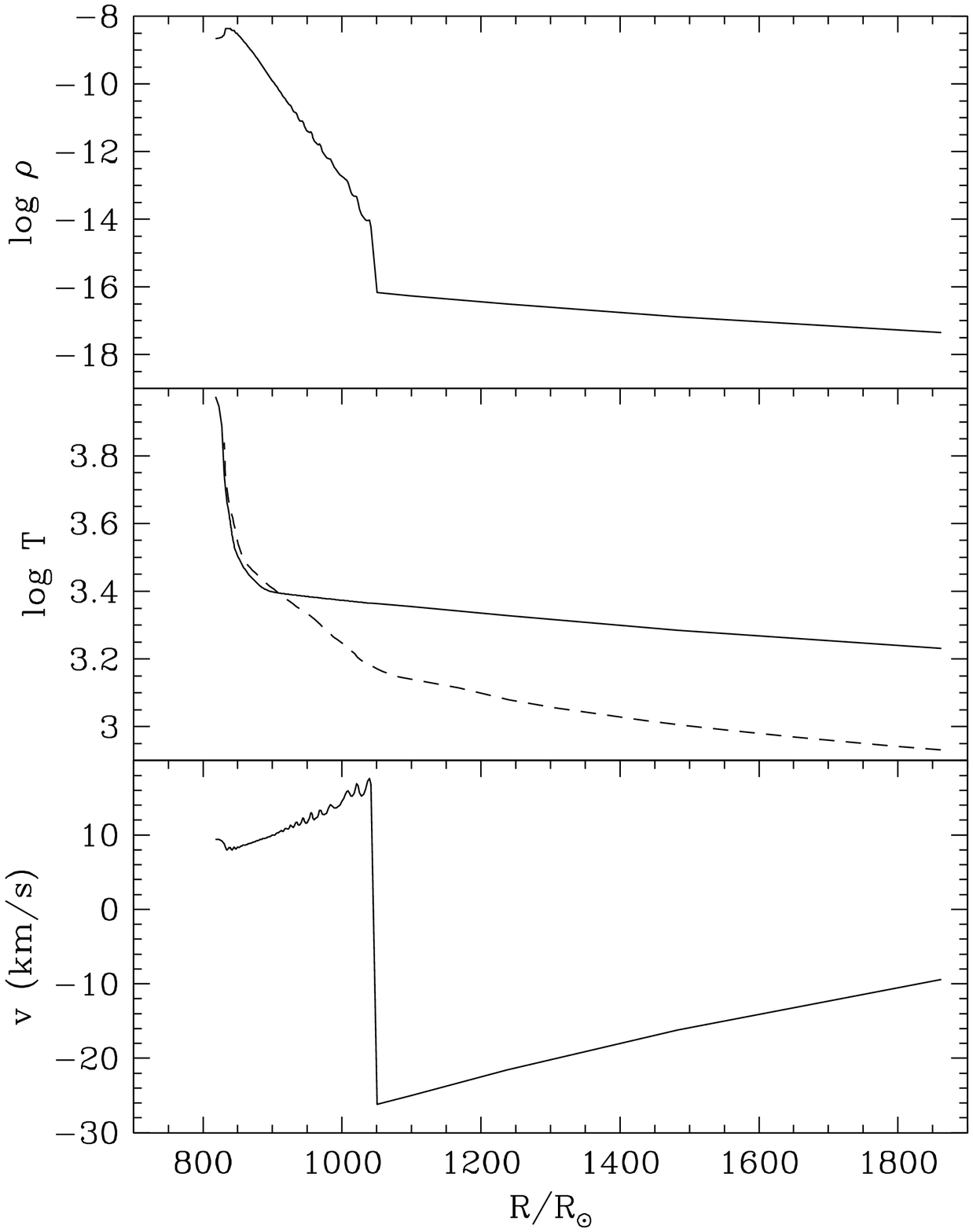}
\caption{The same as Fig.~\ref{fig:1866_atm} but for GM\,103.} 
\label{fig:103_atm}
\end{figure}

\subsection{HV\,2576}

HV\,2576 is an LMC star which is more luminous than NGC\,1866 \#4.  It has a
much longer period, a larger amplitude, and it pulsates in the fundamental
mode.  The derived mass for HV\,2576 is 6 M$_{\odot}$.

Fig.~\ref{fig:2576_lc} shows the observed and computed $K$, $V$ and $M_{\rm bol}$
light curves.  Like many of the intermediate-mass, large-amplitude AGB pulsators in the
Magellanic Clouds, this star shows a double-humped optical light curve.  The
pulsation models do not accurately reproduce the double hump but they do show
some evidence for it.  The $K$ light curve does not show evidence for a
prominent hump, but the points are too sparse in the light curve to be
definitive.  Overall, the period and amplitude of the pulsation and the large
$V$-$K$ colour ($V$-$K$ $\sim$ 7) are reproduced well.  The Gemini observation
was made near minimum light of the $K$ light curve, or between the
double-humped maximum of the optical light curve.  The structure of the
pulsation model at this time is shown in Fig.~\ref{fig:2576_atm}.  The velocity
gradient through the atmosphere at the time of observation was small and there
was no shock wave present in the atmosphere.  As with NGC\,1866 \#4, the
temperature solution from the model atmosphere program produces a cooling of
the outer layers and a warming at moderate optical depths compared to the
modified Eddington approximation used in the pulsation models.

\subsection{HV\,11303}

HV\,11303 is an SMC star which is almost a twin of HV\,2576 in the LMC in terms
of its pulsation period and luminosity.  However, it has a much larger
pulsation amplitude.  It pulsates in the fundamental mode and the derived mass
is 4.6 M$_{\odot}$.  This mass is smaller than that of HV\,2576, which
may explain the larger pulsation amplitude.

Fig.~\ref{fig:11303_lc} shows the observed and computed $K$, $V$ and $M_{\rm
bol}$ light curves.  The pulsation models reproduce the observed $K$ light
curve reasonably well.  The Gemini observation was made near minimum light of
the $K$ light curve.  The structure of the pulsation model at this time is
shown in Fig.~\ref{fig:11303_atm}.  Note that there is an uncertainty of about
0.1 in the phasing of the models relative to the $K$ light curve.  With the
adopted phasing, a strong shock was just emerging into the atmosphere at the
time of Gemini observation.  In addition, there was a large velocity gradient
through the atmosphere above the shock.  The temperature solution from the
model atmosphere program shows a large temperature drop in the outer layers.
This behaviour is typical for extended atmospheres in a certain parameter range
where strong water absorption abruptly dominates in the high layers
\citep{sch85}.  The warmer, grey pulsation model used as input to the atmosphere code
does not have water-forming low temperatures.

\subsection{GM\,103}

GM\,103 (Groenewegen et al. 1995) is also an SMC star but of greater
luminosity, amplitude and period than HV\,11303.  It pulsates in the
fundamental mode and the derived mass is 6 M$_{\odot}$.

Fig. \ref{fig:103_lc} shows the observed and computed $K$, $V$ and $M_{\rm
bol}$ light curves.  The pulsation models reproduce the $K$ light curve and the
large $V$-$K$ colour ($V$-$K$ $\sim$ 8) reasonably well.  However, the $V$
light curve of the pulsation models seems to rise towards maximum earlier than
the observed light curve.  The Gemini observation was made near maximum visible
light, or about phase 0.1 before maximum of the $K$ light curve.  The structure
of the pulsation model at this time is shown in Fig. \ref{fig:103_atm}.  This
star has a strong shock situated in the middle part of the atmosphere at the
time of Gemini observation, with moderate velocity gradients above and below
the shock.  The temperature solution for GM\,103 from the model atmosphere
program shows a larger deviation from the pulsation solution than in any other
star.  This is due to the strong density drop at the shock, with the model
atmosphere's water-dominated temperature dropping rapidly near what is the 
effective edge of the star at the shock front.

\section{Spectral synthesis and abundance derivation}\label{sec:modelatmos}

\subsection{Line synthesis calculations}

The spherical model atmospheres described in Section~\ref{sec:pulsation} were put into a
compatible spectrum synthesis program (using spherical geometry) to generate line spectra.  The spectrum
synthesis program was that used by \citet{sch92}, \citet{bsw96} and \citet{sw00}, except that
partial pressures of a larger number of molecules were calculated using a computer program supplied by
K. Ohnaka (private communication).  This program uses an equation of
state based on molecular constants of \citet{tsuji1973}, updated for CN
using data from \citet{cos90} and for TiO from \citet{tsu78}. Lines are assumed to be
formed in local thermodynamic equilibrium. As shock-heating behind the shock
front is neglected in these models, emission components of
synthetic line profiles may only occur as a consequence of large atmospheric
extension (P-Cygni-like emission; see Scholz 1992).

The overall spectrum in each of the three filter regions was computed using a
reasonably comprehensive set of atomic and diatomic molecular lines, although
only specific lines were used in the actual abundance derivations.  For atomic
lines, wavelengths, excitation potentials and $\log{gf}$ values were taken from
the Vienna Atomic Line Database \citep[VALD,][]{kprsw99}, with parameters of
lines selected for abundance determination being checked against those of
\citet{smith2002}.  Molecular line data proved more problematic as no
single source exists.  CO lines were taken from \citet{gv1994} and CN lines
from \citet{at97} and the SCAN database \citep{jl90}. CH lines were also
obtained from the SCAN database \citep{jliy96}. In each of these cases, lines
were cross-checked against those from \citet{smith2002}. The OH lines proved
most problematic. An incomplete selection of lines was assembled from
\citet{smith2002} and \citet{mbs01}: several additional lines listed in the
source \citet{adreb94} could not be not included due to a lack of $gf$ values.

\subsection{The derivation of abundances}\label{abund_deriv}

Starting abundances were adopted by approximating LMC and SMC metal abundances as
half and quarter solar, respectively, with solar defined from
\citet{asplund2005} (see Table \ref{tab:abundances}). The CO and OH lines
computed with these abundances were examined to establish the microturbulent
velocity.

\begin{figure}
\includegraphics[width=0.45\textwidth]{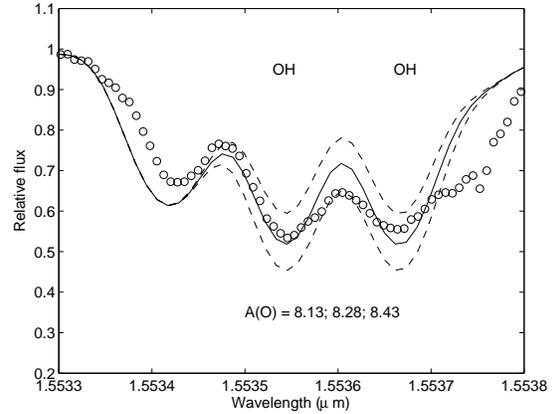}
\caption{Fitted OH lines for HV\,2576.  In this and subsequent figures, fits are shown for the
three abundance values shown on the figure (where $A(X) = \log[n(X)/n(H)] + 12$).  The dashed lines
correspond to changes of $\pm$0.15 dex from the adopted abundance (solid line).}
\label{fig:oheglineshv2576}
\end{figure}

\begin{figure}
\includegraphics[width=0.45\textwidth]{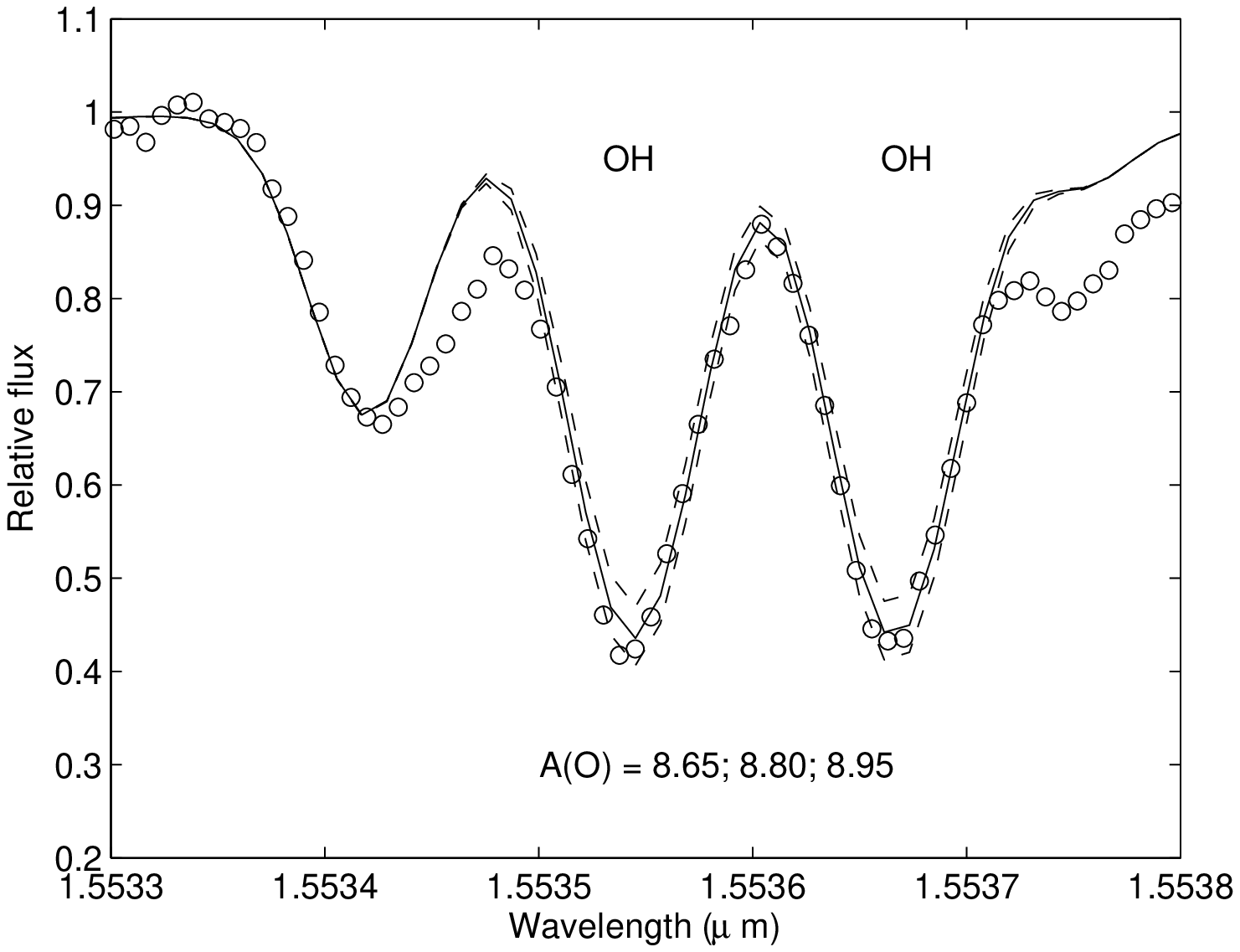}
\caption{Fitted OH lines for NGC\,1866\,\#4.}
\label{fig:oheglinesngc18664}
\end{figure}

\begin{figure}
\includegraphics[width=0.45\textwidth]{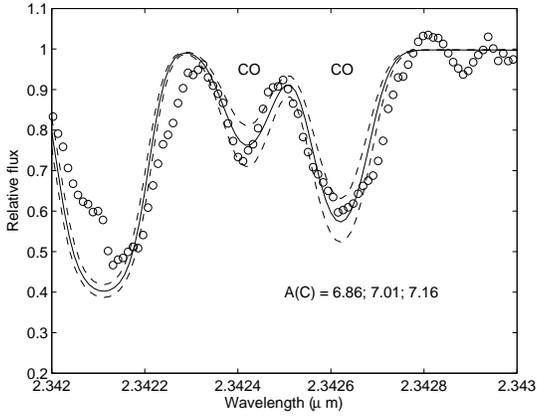}
\caption{Fitted CO lines for HV\,2576.}
\label{fig:coeglineshv2576}
\end{figure}

\begin{figure}
\includegraphics[width=0.45\textwidth]{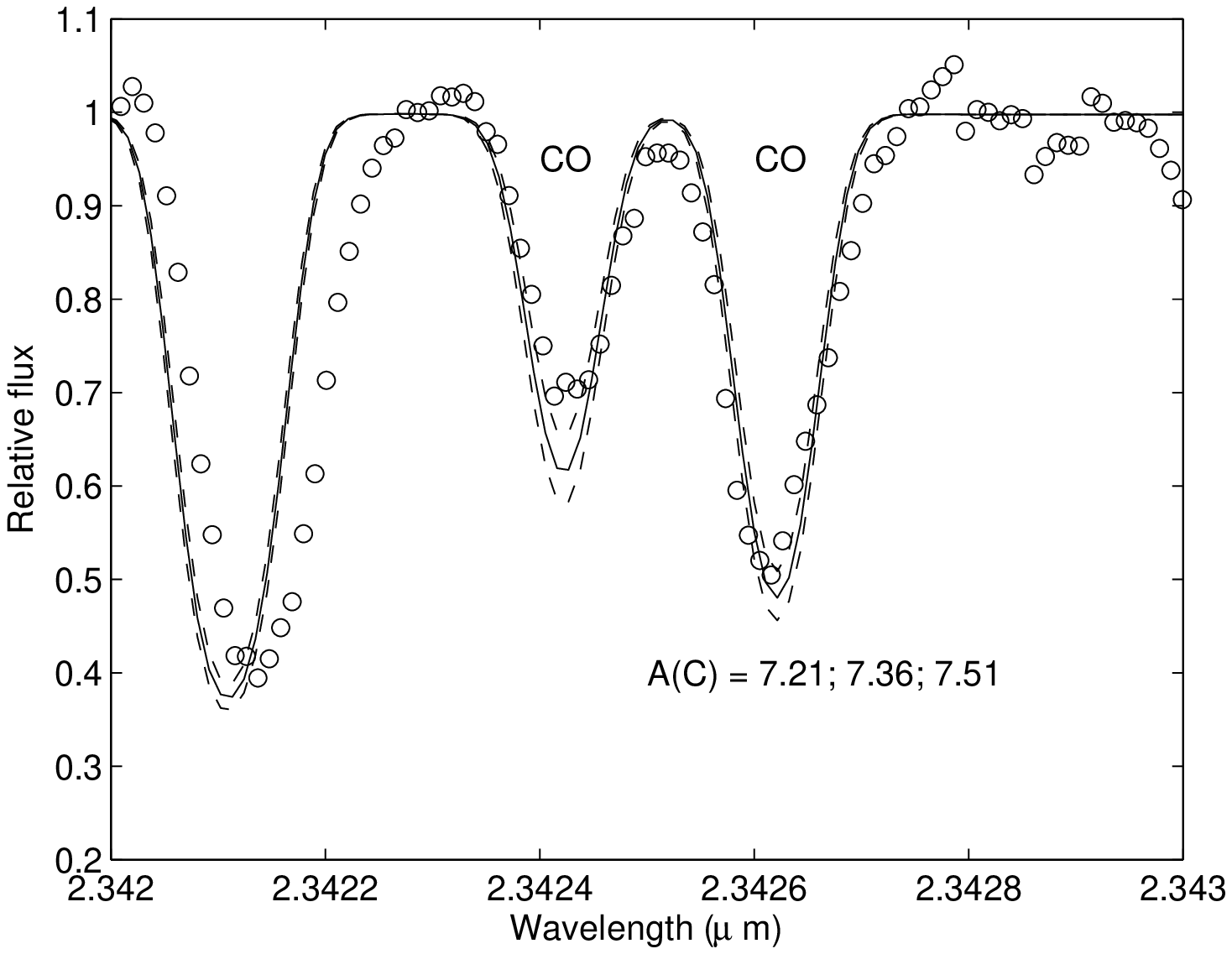}
\caption{Fitted CO lines for NGC\,1866\,\#4.}
\label{fig:coeglinesngc18664}
\end{figure}

\begin{figure}
\includegraphics[width=0.45\textwidth]{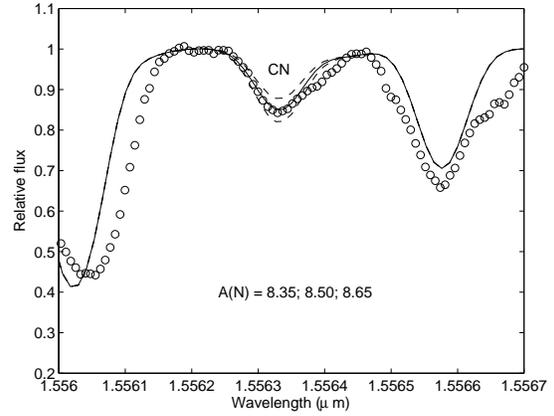}
\caption{Fitted CN line for HV\,2576.}
\label{fig:cneglineshv2576}
\end{figure}

\begin{figure}
\includegraphics[width=0.45\textwidth]{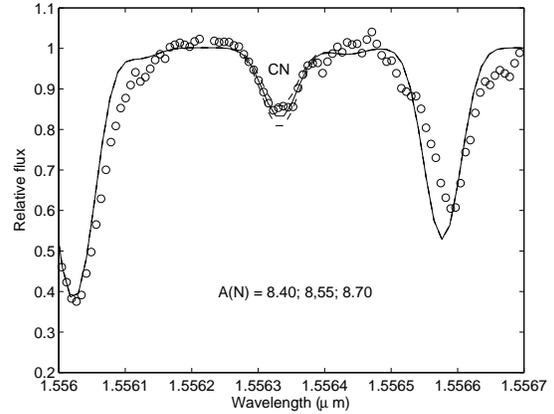}
\caption{Fitted CN line for NGC\,1866\,\#4.}
\label{fig:cneglinesngc18664}
\end{figure}

\begin{figure}
\includegraphics[width=0.45\textwidth]{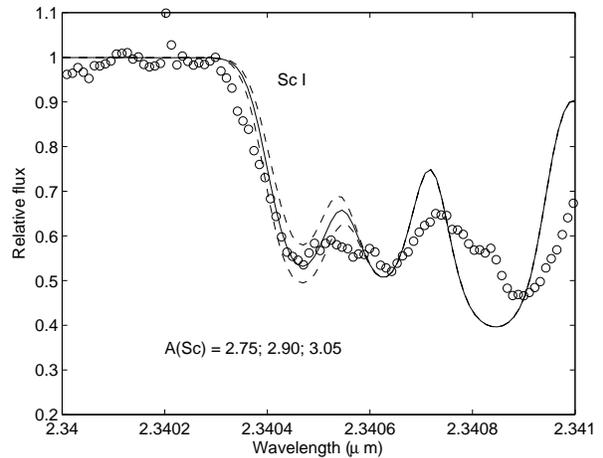}
\caption{Fitted Sc I line for HV\,2576.}
\label{fig:sceglineshv2576}
\end{figure}

\begin{figure}
\includegraphics[width=0.45\textwidth]{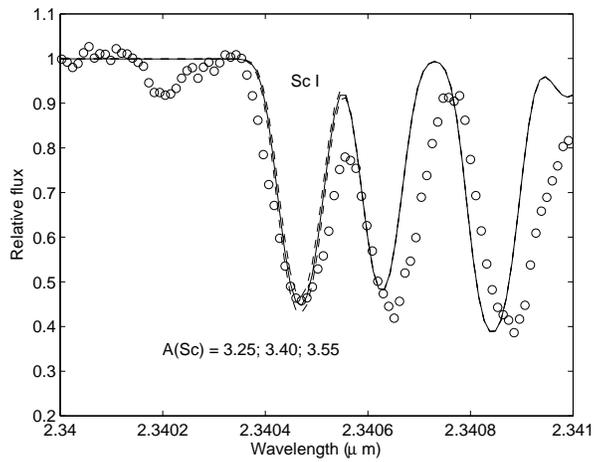}
\caption{Fitted Sc I line for NGC\,1866\,\#4.}
\label{fig:sceglinesngc18664}
\end{figure}

\begin{figure}
\includegraphics[width=0.45\textwidth]{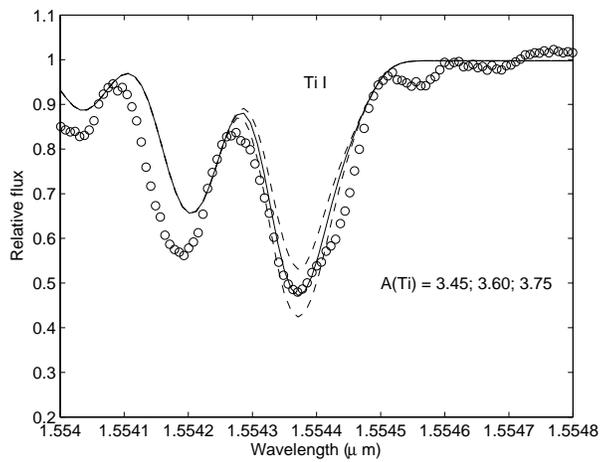}
\caption{Fitted Ti I line for HV\,2576.}
\label{fig:tieglineshv2576}
\end{figure}

\begin{figure}
\includegraphics[width=0.45\textwidth]{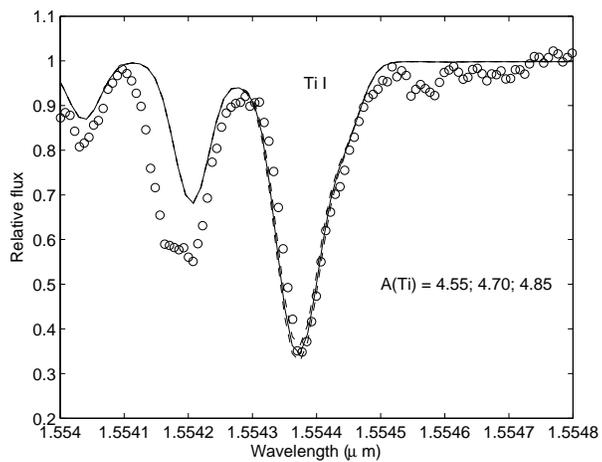}
\caption{Fitted Ti I line for NGC\,1866\,\#4.}
\label{fig:tieglinesngc18664}
\end{figure}

\begin{figure}
\includegraphics[width=0.45\textwidth]{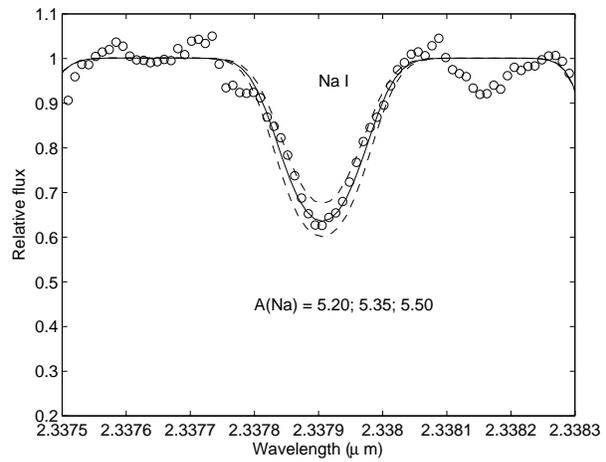}
\caption{Fitted Na I line for HV\,2576.}
\label{fig:naeglineshv2576}
\end{figure}

\begin{figure}
\includegraphics[width=0.45\textwidth]{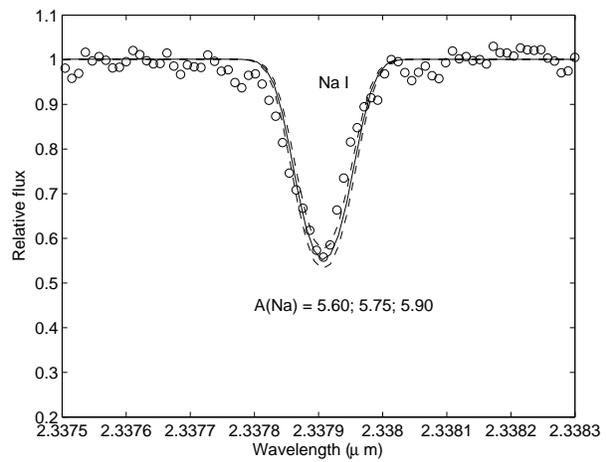}
\caption{Fitted Na I line for NGC\,1866\,\#4.}
\label{fig:naeglinesngc18664}
\end{figure}

\begin{figure}
\includegraphics[width=0.45\textwidth]{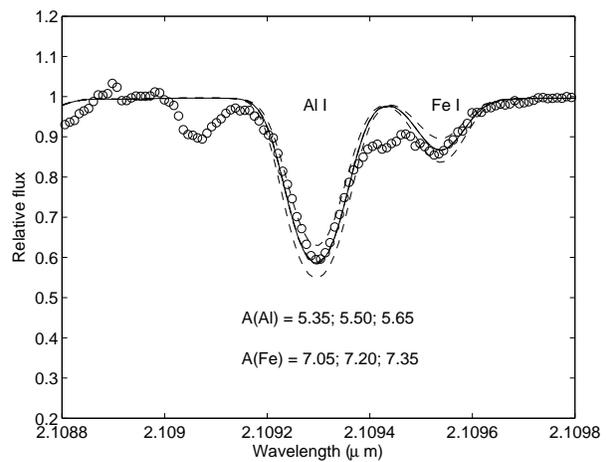}
\caption{Fitted Fe I and Al lines for HV\,2576.}
\label{fig:alfeeglineshv2576}
\end{figure}

\begin{figure}
\includegraphics[width=0.45\textwidth]{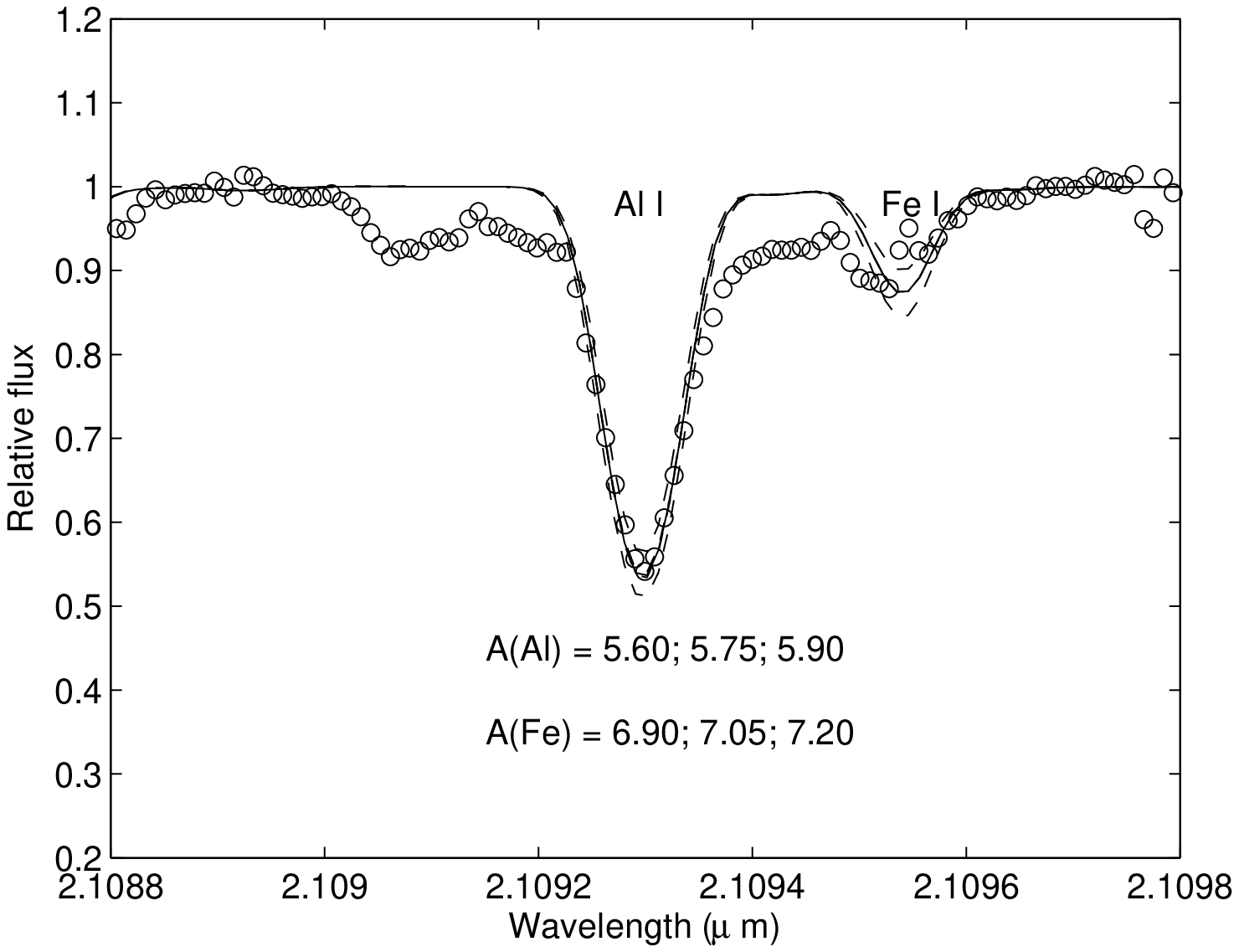}
\caption{Fitted Fe I and Al lines for NGC\,1866\,\#4.}
\label{fig:alfeeglinesngc18664}
\end{figure}

As these relatively massive stars are expected to have altered CNO abundances
as a result of nucleosynthetic processing, a series of CNO values based on the
AGB evolution calculations of \citet{ak03} and \citet{kl03} was adopted, and
the resultant spectra were then calculated.  The CNO combination giving the
best overall fit to the observed spectra was used as the primary starting point
for subsequent refinement of the individual CNO abundances.  Once the best CNO
abundances had been found, the metallic abundances were refined to give a best
fit.  Lines used for final abundance derivations are listed in Table
\ref{tab:spectrallines}, and a selection of the fitted lines are shown in
Figs.~\ref{fig:oheglineshv2576} to \ref{fig:alfeeglinesngc18664}.

The abundances derived from different lines of the same species show some
scatter, with the adopted lines generally found to yield reasonably consistent
values.  Figs.~\ref{fig:coeglineshv2576} and \ref{fig:coeglinesngc18664} show
examples of the fit errors for the CO lines.  It is clear that neither CO line
is fitted exactly by the final adopted abundance (that of the mean of the best
fits to 6 lines), but both lines are fitted reasonably well, with one too
strong and one too weak.  Figs.~\ref{fig:oheglineshv2576} and
\ref{fig:oheglinesngc18664}, where Fe I lines are present on either side of the
OH lines, show further examples of fit errors.  The adopted Fe abundance is
based on a fit to the 2.1095$\mu$m line in Figs.~\ref{fig:alfeeglineshv2576}
and \ref{fig:alfeeglinesngc18664}.  However, in HV\,2576
(Fig.~\ref{fig:oheglineshv2576}), this abundance is too high for the
1.553424$\mu$m Fe line to the blue and too low for the 1.553757$\mu$m line to
the red (or, more likely, the models are missing an unidentified line at this
wavelength).  For NGC\,1866\,\#4 (Fig.~\ref{fig:oheglinesngc18664}), the
1.553424$\mu$m line fits well, while once again the 1.553757$\mu$m line is too
weak (or an unidentified line is missing in the model).

There were also cases where observed and computed line velocities do not match
to better than a few km s$^{-1}$, as can be seen in
Figs.~\ref{fig:coeglineshv2576} to \ref{fig:cneglinesngc18664}.  The velocity
mismatch could be due to a number of potential causes, including inexact
adopted line wavelengths, inexact wavelength calibration, or the model
velocities as a function of depth not being correct.

The microturbulent velocity required for line broadening varied from star to
star.  The adopted value for NGC\,1866\,\#4 (3 km s$^{-1}$) is close to the value
used by \citet{smith2002}.  However, a larger value of 7 km s$^{-1}$ was
required for HV\,2576.  This value was derived by fitting to the CO lines.
Slightly different values would be obtained by fitting different lines.  This
is probably a reflection of different amounts of turbulence at different levels
in the atmosphere: the CO lines tend to be broader than the metallic lines and
they are, on average, formed further out in cooler layers.

\begin{table}
\caption{Spectral line data.}
\label{tab:spectrallines}
\begin{tabular}{lccl}
\hline
$\lambda$ & $\chi$ & & Source \\
(\AA) & (eV) & $\log(gf)$ & \\
\hline
Fe I \\
21095.446 & 6.20 & -0.69 & VALD \\
Na I \\
23378.945 & 3.75 & -0.420 & \citet{smith2002} \\
Sc I \\
23404.756 & 1.44 & -1.278 & \citet{smith2002} \\
Ti I \\
15543.720 & 1.88 & -1.481 & \citet{smith2002} \\
Al I \\
21093.029 & 4.09 & -0.31 & VALD \\
OH \\
15535.489 & 0.51 & -5.23 & \citet{smith2002} \\
15536.707 & 0.51 & -5.23 & \citet{mbs01} \\
15560.271 & 0.30 & -5.31 & \citet{smith2002} \\
15565.815 & 0.90 & -5.00 & \citet{mbs01} \\
CO \\
23396.260 & 0.37 & -5.17 & \citet{smith2002} \\
23398.224 & 1.73 & -4.44 & \citet{smith2002} \\
23406.345 & 0.00 & -6.57 & \citet{smith2002} \\
23408.509 & 0.36 & -5.19 & \citet{smith2002} \\
23424.328 & 1.80 & -4.427 & \citet{smith2002} \\
23426.322 & 0.00 & -6.565 & \citet{smith2002} \\
CN\\
15563.367 & 1.15 & -1.141 & \citet{smith2002} \\
\hline
\end{tabular}
\end{table}

Our final derived abundances are shown in Table \ref{tab:abundances}.
For the elements Fe, Na, Al, Sc, Ti, and N, the
error estimates in the table are those arising from the fit to the
line only, as only one line was available.
They are eye-estimates based on the fits 0.15 dex above and below the
adopted optimum fit abundance.  The C and O values are based on
averages of the fits to 6 and 4 lines, respectively, with the stated
errors being based on the standard deviation. The lines used are given
in Table \ref{tab:spectrallines}. Systematic errors, such as errors in
the pulsation model and its phase, will undoubtedly add significantly
to the fit error, but such additional errors are hard to estimate.

\begin{figure*}
\includegraphics[width=\textwidth]{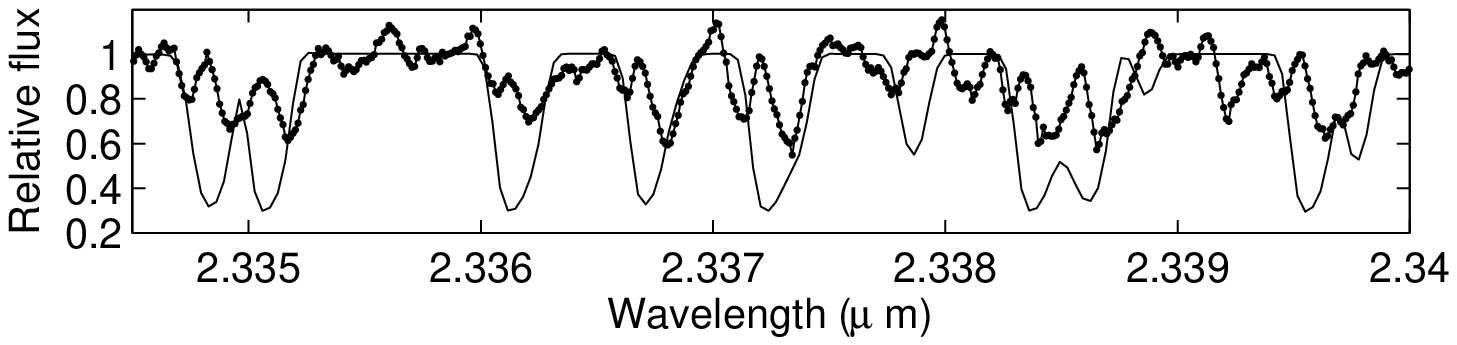}
\includegraphics[width=\textwidth]{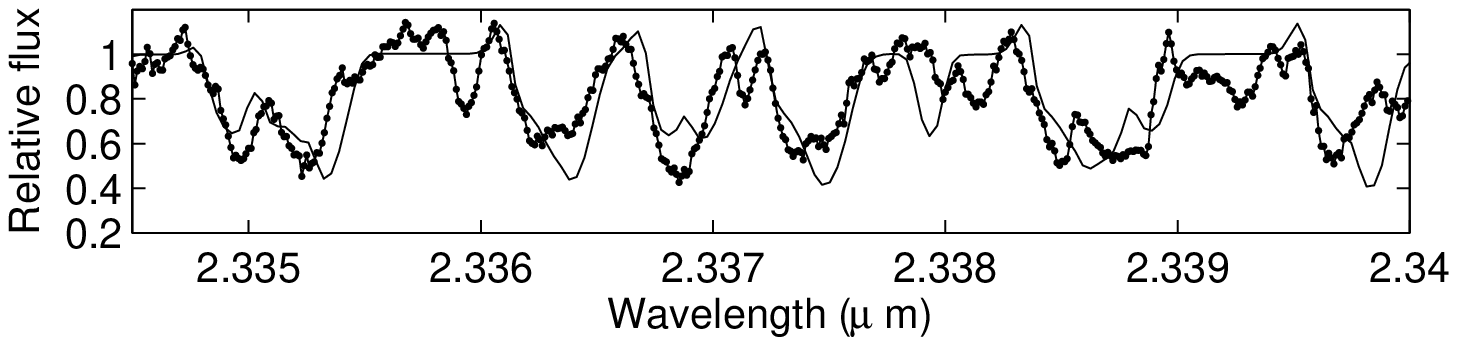}
\caption{Top: A section of the 2.34 $\mu$m
spectrum of GM\,103 (points), overlayed with a synthesised spectrum (continuous line).
Bottom: The same piece of spectrum for HV\,11303.
}
\label{fig:gm103_M119467_k4308}
\end{figure*}

\begin{table*}
\caption{Derived abundances}
\label{tab:abundances}
\begin{tabular}{lcccccccc}
\hline
Star & A(Fe) & A($^{12}$C) & A($^{14}$N) & A($^{16}$O) & A(Na) & A(Sc) & A(Ti) & A(Al) \\
\hline
\multicolumn{9}{l}{LMC}\\
HV 2576 & 7.20 $\pm$ 0.10
        & 7.01 $\pm$ 0.20 
        & 8.50 $\pm$ 0.05 
        & 8.28 $\pm$ 0.08 
        & 5.35 $\pm$ 0.10 
        & 2.90 $\pm$ 0.10 
        & 3.60 $\pm$ 0.05 
        & 5.50 $\pm$ 0.05
\\
NGC 1866 \#4    & 7.05 $\pm$ 0.15
                & 7.36 $\pm$ 0.16 
                & 8.55 $\pm$ 0.15 
                & 8.80 $\pm$ 0.17 
                & 5.75 $\pm$ 0.15
                & 3.40 $\pm$ 0.15 
                & 4.70 $\pm$ 0.15 
                & 5.75 $\pm$ 0.10
\\
Smith et al. 2002  & 6.37 - 7.16 & 6.53 - 7.86 & 7.14 - 8.24 & 7.82 - 8.33 & 4.69 - 5.84 & 2.01 - 2.91 & 3.84 - 4.61 & ... \\
\multicolumn{9}{l}{}\\
Half-solar              & 7.15
                        & 8.09
                        & 7.48
                        & 8.36
                        & 5.87
                        & 2.75
                        & 4.60
                        & 5.87
\\
\hline
\end{tabular}
\begin{flushleft}
\footnotesize{Notes: $A(X) = \log[n(X)/n(H)] + 12$.  Half-solar abundances are half the solar
abundances in \citet{asplund2005}.  The errors are estimates of the line fitting error 
alone - see Sect.~\ref{abund_deriv}.}
\end{flushleft}
\end{table*}

\subsection{HV\,11303 and GM\,103}

As noted in Section~\ref{sec:pulsation}, the SMC stars HV\,11303 and GM\,103 both had
strong shock fronts in their atmospheres at the time of observation with Gemini South.
This has made abundance derivation with our atmosphere code impossible.  Fig.~\ref{fig:gm103_M119467_k4308}
shows a sample of the spectrum for each star, along with an attempt at synthesisizing
the spectrum using the pulsation models from Section~\ref{sec:pulsation}.

In the case of GM\,103, the synthesized absorption lines are much stronger than
the observed lines.  A comparison of the observed and synthesisized spectra
shows clearly that the observed line cores are filled in by strong emission in
every case.  An alternative explanation could be that each line is split into a
pre- and post-shock absorption component, with no emission in the middle.
However, the weakness of the observed absorption lines compared to the model
lines suggests that there is indeed a strong emission component.  In HV\,11303,
the lines are much broader than in GM\,103, with evidence for an emission
component as well as absorption components corresponding to the pre- and
post-shock velocities.

As noted in Section~\ref{sec:pulsation}, the spectral synthesis code ignores
the immediate post-shock, high-temperature region so that there is no
post-shock emission component in our synthesized spectra.  When a strong shock
is present in the atmosphere, this model deficiency prevents abundance
determination from lines which have a post-shock emission contribution.

These results show that observations useful for abundance analysis must be
obtained at phases of the pulsation cycle when no strong shock wave is present
in the atmosphere.  In these relatively massive stars, the shock wave enters
the lower part of the stellar atmosphere near minimum of the $K$ light curve, or
phase $\sim$0.2--0.3 past minimum visual light.  It has passed through the line
forming part of the atmosphere half a pulsation cycle later, corresponding to
$\sim$0.1--0.2 in phase past $K$ maximum or after phase $\sim$0.4 of the visual
light curve.

\section{Discussion}

\begin{figure}
\includegraphics[bb=100 250 525 553, width=0.5\textwidth]{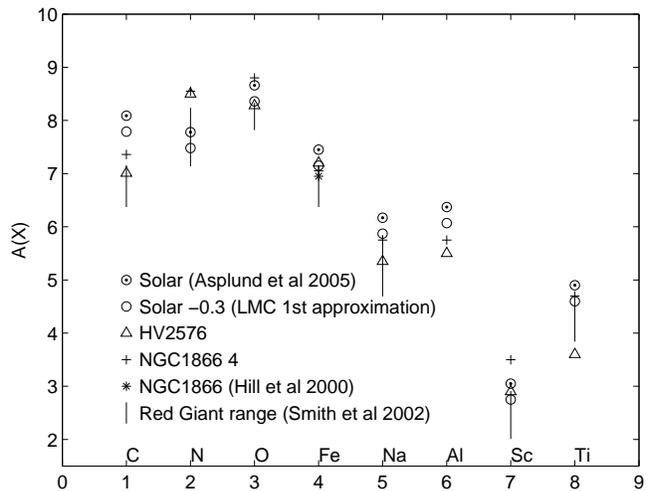}
\caption{Final abundances.}
\label{fig:finalabundances}
\end{figure}

The abundances obtained here for NGC\,1866\,\#4 and HV\,2576 are shown with
other comparable observations in Fig.~\ref{fig:finalabundances}.  The Fe
abundance is close to half solar, consistent with the values obtained by
\citet{smith2002}, \citet{hfsps00} and other studies of young objects in the
LMC.  The other metal abundances are similarly near half-solar, except for Ti
in HV\,2576 where our derived value seems anomalously low.

The CNO abundances are the most interesting part of this study, since in a
highly-evolved, intermediate-mass AGB star they will be altered by the 1st, 2nd and 3rd
dredge-ups and by hot-bottom burning.

Evidence for hot-bottom burning has been found in intermediate-mass AGB stars in the Magellanic Clouds
by \citet{psl93} and \citet{smith1995}.  \citet{psl93} examined spectra of 7
luminous AGB stars in the SMC and found them to be Li-rich, C-poor and with low
$^{12}$C/$^{13}$C ratios, consistent with hot-bottom burning.  \citet{smith1995} examined a
large number of SMC and LMC stars and found many luminous AGB stars, including
HV\,2576, showed strong Li lines that could be attributed to hot-bottom burning.  In fact,
HV\,2576 had the largest estimated Li abundance of all the stars examined.
\citet{mac02} examined Li lines in the giant stars in NGC\,1866 and found that
NGC\,1866\,\#4 (which they mislabel as \#2) had enhanced Li compared to other
stars in the cluster, indicating that hot-bottom burning is operating in this star also.

There is evidence that the third dredge-up has been operating in HV\,2576, as
well as hot-bottom burning.  \citet{wbf83} give a spectral type M5.5/S1 for
HV\,2576, indicating the presence of excess ZrO in the spectrum, which is a
result of the dredge-up of s-process elements during thermal pulses.
\citet{smith1995} noted that HV\,2576 had one of the strongest ZrO bands in
their large sample of stars, indicating that a large amount of third dredge-up
has occurred in this star.

The combination of third dredge-up and hot-bottom burning is predicted to be a
significant source of primary nitrogen in the universe.  The $^{12}$C dredged
up at a thermal pulse is converted into $^{14}$N during the next interpulse
period when hot-bottom burning operates during the quiescent H-shell burning
phase.  \citet{lw04} and \citet{ak03} (see their Figs. 2.37 and 5.6,
respectively) show the effects of third dredge-up and hot-bottom burning in
intermediate-mass, luminous and sub-solar metallicity AGB stars such as those we are
dealing with here.  Initially on the AGB, just at the onset of thermal pulses,
hot-bottom burning by the CN cycle reduces the $^{12}$C abundance to $\sim$1/15
of its initial main-sequence value, and increases the $^{14}$N abundance to
$\sim$5-6 times its initial value.  The ratio $^{12}$C/$^{14}$N at this time is
$\sim$1/15.  During subsequent evolution, the dredge-up of $^{12}$C by third
dredge-up at thermal pulses, followed by hot-bottom burning, causes a steady
rise over many pulsation cycles in both the C and N abundance, with the C/N
ratio remaining almost constant at $\sim$1/15.  During this relatively long
evolutionary process, there is a small decrease in the O abundance, but this
effect would be too small for us to reliably detect.

Looking at Fig.~\ref{fig:finalabundances} and Table~\ref{tab:abundances}, we
see that the C abundance in both NGC\,1866\,\#4 and HV\,2576 is about 0.1 times
half-solar.  At the same time, the N abundance is $\sim$10 times the half-solar
value.  Both these numbers are consistent with the strong operation of
hot-bottom burning in the envelopes of these stars.  Furthermore, the very
large N abundances can only be explained by hot-bottom processing of dredged-up
$^{12}$C into $^{14}$N i.e.  a large fraction of the nitrogen in the envelopes
of these stars must be of primary origin.  This is the first direct
demonstration that this long-predicted source of primary nitrogen does exist.

The O abundance in HV\,2576 is essentially unchanged from the half-solar value,
as predicted by models.  There is some evidence for an increase in the O
abundance in NGC\,1866\,\#4.  If confirmed, this would lend support to
convective theories such as that of \citet{her97} which lead to overshoot at
convective boundaries, enrichment of the intershell region in both $^{12}$C and
$^{16}$O at each helium shell flash, and subsequent dredge-up of both $^{12}$C and
$^{16}$O in third dredge-up events.

A strong disagreement between the observations and evolutionary models without
convective overshoot relates to the total stellar mass required for the
efficient operation of hot-bottom burning.  The mass of NGC\,1866\,\#4 must be
no more than $\sim$4 M$_{\odot}$ because of cluster membership.  This star shows evidence
for efficient hot-bottom burning, while models without convective overshoot do
not predict efficient hot-bottom burning at masses this low.  This suggests
that overshoot (inwards) of convective material at
convective boundaries is required to significantly extends the convective
envelope into nuclear-burning regions.  The models of \citet{ven02} use a
convective theory that does produce overshoot, and their 4 M$_{\odot}$,
half-solar metal abundance AGB models do have the efficient hot-bottom burning observed in
NGC\,1866\,\#4.  At 6 M$_{\odot}$, both the observations of HV\,2576 and models
with or without convective overshoot show efficient hot-bottom burning.

The elements Li, Na and Al can also be compared with evolutionary models.
\citet{smith1995} give A(Li) = 3.8 for HV\,2576 while \citet{mac02} give A(Li)
= 1.5 (note that these authors used static model atmospheres, and observations
at random phases where shocks may have been present in the stellar atmosphere).  
Models of AGB stars with efficient hot-bottom burning generally show an
initial large increase in the surface Li abundance, after which it slowly decreases to low
values (A(Li) $\la$ 1.0) while the CN cycle converts C into N -
see \citet{ak03} and \citet{ven02}.  The high Li value derived for HV\,2576, together with
the observed high $^{14}$N and low $^{12}$C abundances, is consistent 
with model predictions for a 6 M$_{\odot}$ star about one quarter of the way through
its TPAGB phase and undergoing efficient hot-bottom burning - see Fig. C27 of \citet{ak03}.

The Na and Al abundances found in this study are marginally below the expected
half-solar values and within the expected range of observed variation.  Models
of AGB stars \citep{ak03} with efficient hot-bottom burning show a small
increase in Al abundance and small changes in the total Mg abundance but these
are too small to detect here.  The $^{25}$Mg/$^{24}$Mg ratio can increase
greatly in such stars, but we are unable to detect isotopic changes.

\section{Summary}

High-dispersion, near-infrared spectra have been used to derive abundances of
luminous, intermediate-mass AGB stars in the Magellanic Clouds.  The C and N
abundances derived here provide the first observational demonstration that
third dredge-up at thermal pulses followed by hot-bottom burning can produce
significant amounts of primary nitrogen in intermediate-mass AGB stars.  The
observed spectra and modelling show that only spectra taken during the part of
the pulsation cycle near minimum visual light, equivalent to the latter part of
the decline of the $K$ light curve, are useful for abundance analysis.

The results in this paper come from exploratory observations and calculations
and, in the end, only two stars were suitable for abundance analysis.  It is
clear that follow-up work with a larger sample of objects is required to
confirm the results found here, and to seek C and N abundances as a function of
stellar mass, luminosity and metal abundance.

\section*{Acknowledgments}

We would like to thank Amanda Karakas for useful discussions about AGB
nucleosynthesis models, Patrick Tisserand for supplying the Eros2 light curve
for HV 11303, and Keiichi Ohnaka for supplying his programs for partial
pressure calculations.  JAM is grateful for the discovery grant from the
Australian Research Council (DP0343832) which supported her during the course
of this work: PRW and JCL also received partial support from this grant.  MS
received support from a grant of the Deutsche Forschungsgemeinschaft.  This
paper is based in part on observations obtained at the Gemini Observatory,
which is operated by the Association of Universities for Research in Astronomy,
Inc., under a cooperative agreement with the NSF on behalf of the Gemini
partnership: the National Science Foundation (United States), the Particle
Physics and Astronomy Research Council (United Kingdom), the National Research
Council (Canada), CONICYT (Chile), the Australian Research Council (Australia),
CNPq (Brazil), and CONICRT (Argentina).  The observations were obtained with
the Phoenix infrared spectrograph, which was developed and is operated by the
National Optical Astronomy Observatory.  The spectra were obtained as part of
Gemini programs GS-2002A-Q-49, GS-2002B-Q-22 and GS-2003A-Q-25.  We thank the
observers at Gemini South for taking these spectra in service mode.

\bsp

\label{lastpage}

\end{document}